\documentclass[prd,twocolumn,groupedaddress,showpacs,nofootinbib]{revtex4}
\usepackage{graphicx}
\usepackage{dcolumn}
\usepackage{amssymb}
\usepackage{mathrsfs}
\usepackage{amsmath}
\usepackage{epsfig}
\usepackage[dvips]{color}
\usepackage{hhline}

\begin{document}

\title{Common origin of inverse seesaw and baryon asymmetry}

\author{Pei-Hong Gu}

\email{peihong.gu@sjtu.edu.cn}

\affiliation{School of Physics and Astronomy, Shanghai Jiao Tong University, 800 Dongchuan Road, Shanghai 200240, China}

\begin{abstract}

In the inverse seesaw scenario, several fermion singlets have a small Majorana mass term. We show such Majorana masses can be suppressed by some heavy fermion and/or Higgs singlets after a global symmetry is spontaneously broken. These interactions can also accommodate a leptogenesis mechanism to explain the cosmic baryon asymmetry.

\end{abstract}


\maketitle

\section{Introduction}

The phenomenon of neutrino oscillations has been established by the atmospheric, solar, accelerator and reactor neutrino measurements. This fact implies three flavors of neutrinos should be massive and mixed \cite{pdg2018}. Moreover, the cosmological observations indicate that the neutrinos should be extremely light \cite{pdg2018}. The tiny but nonzero neutrino masses call for new physics beyond the standard model (SM). The seesaw \cite{minkowski1977} mechanism is considered the best explanation for the smallness of the neutrino masses. Some seesaw models \cite{minkowski1977,mw1980} can also accommodate a leptogenesis \cite{fy1986,lpy1986,fps1995,ms1998,bcst1999,hambye2001,di2002,gnrrs2003,hs2004,bbp2005} mechanism to generate the cosmic baryon asymmetry, which is another big challenge to the SM.

The inverse \cite{mv1986} seesaw has become one of the  most attractive seesaw scenarios because of its testability. In the inverse seesaw models, several neutral fermions have a small Majorana mass term and mix with the same number of right-handed neutrinos. Due to such small Majorana masses, the right-handed neutrinos can have a sizable Dirac mass term with the left-handed neutrinos. However, the testable inverse seesaw has few disadvantages. For example, it probably needs an additional origin for the small Majorana masses of the neutral fermions. It also seems to have no ideas for the generation of the baryon asymmetry.

In this paper we shall extend the SM $SU(3)_c^{}\times SU(2)_L^{}\times U(1)^{}_{Y}$ gauge symmetries by a $U(1)_{B-L}^{}$ gauge symmetry and a $U(1)_X^{}$ global symmetry. After a gauge-singlet Higgs scalar develops its vacuum expectation value (VEV) for spontaneously breaking the $U(1)_X^{}$ symmetry, three gauge-singlet fermions can obtain their small Majorana masses by integrating out some heavy gauge-singlet fields. These gauge-singlet fermions can also have a mass term with three right-handed neutrinos after the $U(1)_{B-L}^{}$ symmetry is spontaneously broken. Due to the Yukawa couplings of the right-handed neutrinos to the SM lepton and Higgs doublets, we eventually can realize an inverse seesaw mechanism for generating the left-handed neutrino masses. On the other hand, the heavy gauge-singlet fields can decay to produce an asymmetry stored in the three gauge-singlet fermions. The sphaleron processes \cite{krs1985} then can partially transfer this asymmetry to a baryon asymmetry because of the sizable Yukawa couplings involving the right-handed neutrinos.

\section{The model}

Besides the SM fermions, 
\begin{eqnarray}
\!\!\!\!\!\!&&\begin{array}{l}q^{}_{L}(3,2,+\frac{1}{6},+\frac{1}{3})\,,\end{array} \begin{array}{l}d^{}_{R}(3,1,-\frac{1}{3},+\frac{1}{3})\,,\end{array}  \begin{array}{l}u^{}_{R}(3,1,+\frac{2}{3},+\frac{1}{3})\,,\end{array} \nonumber\\
\!\!\!\!\!\!&&\begin{array}{l}l^{}_{L}(1,2,-\frac{1}{2},-1)\,,\end{array}\begin{array}{l}e^{}_{R}(1,1,-1,-1)\,,\end{array} 
\end{eqnarray}
our model contains three right-handed neutrinos and two types of gauge-singlet fermions,
\begin{eqnarray}
&&\begin{array}{l}\nu^{}_{Ri}(1,1,0,-1)\,,\end{array}
\begin{array}{l}S^{}_{Ri}(1,1,0,0)\,,\end{array} \begin{array}{l}X^{}_{Rb}(1,1,0,0)\,,\end{array}\nonumber\\
&&(i=1,2,3; b=1,...)\,.\end{eqnarray}
In the scalar sector, there are three types of Higgs singlets,
\begin{eqnarray}
\!\!\!\!\begin{array}{l}\chi(1,1,0,-1)\,,\end{array}
\begin{array}{l}\xi(1,1,0,0)\,,\end{array}
\begin{array}{l}\Sigma_a^{}(1,1,0,0)\,,~(a=1,...)\,,\end{array}
\end{eqnarray}
in addition to the SM Higgs doublet,
\begin{eqnarray}
\begin{array}{l}\phi^{}(1,2,-\frac{1}{2},0)\,.\end{array}
\end{eqnarray}
In the above, the brackets following the fields describe the transformations under the $SU(3)_c^{} \times SU(2)^{}_{L}\times U(1)_Y^{} \times U(1)_{B-L}^{}$ gauge groups. We also impose a $U(1)_X^{}$ global symmetry under which the fermions and scalars carry the $X$-number as below,
\begin{eqnarray}
\!\!U(1)_X^{}:&&\!\!\!\!\!\!\!\!\begin{array}{l}q_L^{}(-\frac{1}{3})\,,~d_R^{}(-\frac{1}{3})\,,~u_R^{}(-\frac{1}{3})\,,~l_L^{}(+1)\,,~e_R^{}(+1)\,,\end{array}\nonumber\\
\!\!&&\!\!\!\!\!\!\!\!\begin{array}{l}\nu_{R}^{}(+1)\,,~S_{R}^{}(-1)\,,~X_R^{}(0)\,,~\phi(0)\,,~\chi(0)\,,\end{array}\nonumber\\
\!\!&&\!\!\!\!\!\!\!\!\begin{array}{l}\xi(-1)\,,~\Sigma(+2)\,.\end{array}\end{eqnarray}

For simplicity, we do not write down the full Lagrangian. Instead, we only give the terms relevant to our demonstrations, 
\begin{eqnarray}
\label{lag}
\mathcal{L}&\supset& -\left(M_\Sigma^2 \Sigma^\dagger_{}\Sigma + \rho_\Sigma^{}\Sigma \xi^2_{} + \frac{1}{2}g_\Sigma^{}\Sigma \bar{S}_R^{c} S_R^{} +\textrm{H.c.}\right)\nonumber\\
&&-\left(\frac{1}{2}M_X^{} \bar{X}_R^{c}X_R^{} +g_X^{} \xi \bar{S}_R^{} X_R^{c} +\textrm{H.c.}\right) \nonumber\\
&&- \left(y \bar{l}_{L}^{} \phi \nu_R^{} + f \chi \bar{\nu}_R^{} S_R^c +\textrm{H.c.}\right)\,,
\end{eqnarray}
where the Majorana mass matrix $M_X^{}$ and the Yukawa couplings $g_\Sigma^{}$ are symmetric, i.e. $M_X^{}=M_X^T$, $g_\Sigma^{} = g_\Sigma^T$. Without loss of generality and for convenience, we will work in the basis where the cubic couplings $\rho_\Sigma^{}$ are real while the mass matrices $M_\Sigma^{}$ and $M_{X}^{}$ are real and diagonal. Accordingly, we can define the Majorana fermions as below,
\begin{eqnarray}
X_b^{}= X_{Rb}^{} + X_{Rb}^c = X^c_{b}\,.
\end{eqnarray}
Clearly, the $U(1)_X^{}$ global symmetry is exactly conserved so that the gauge-singlet fermions $S_R^{}$ can not have a gauge-invariant Majorana mass term. Note the $U(1)_X^{}$ symmetry can be identified to a $U(1)_{L-B}^{}$ global symmetry. Alternatively, we can consider other global symmetry under which the $U(1)_{B-L}^{}$ Higgs singlet $\chi$ is also non-trivial. In this case, the global lepton number is allowed to be exactly conserved or softly broken in Eq. (\ref{lag}).

\section{Inverse seesaw}

\begin{figure*}
\vspace{6cm} \epsfig{file=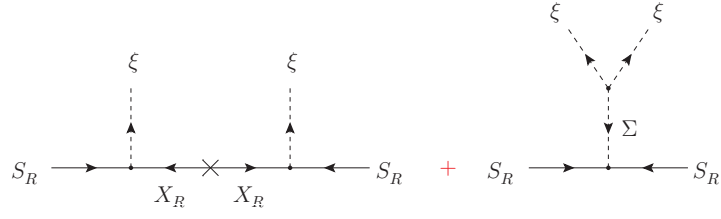, bbllx=5.5cm, bblly=6.0cm,
bburx=15.5cm, bbury=16cm, width=6cm, height=6cm, angle=0,
clip=0} \vspace{-8cm} \caption{\label{numass} The Majorana mass generation of the gauge-singlet fermions.}
\end{figure*}

After the Higgs singlet $\xi$ develops its VEV for spontaneously breaking the $U(1)_X^{}$ symmetry, the heavy Higgs singlets $\Sigma$ can pick up their suppressed VEVs,
\begin{eqnarray}
\langle\Sigma\rangle\simeq - \frac{\rho_{\Sigma}^{} \langle\xi\rangle^2_{}}{M_{\Sigma}^2} \ll \langle \xi\rangle ~~\textrm{for}~~M_\Sigma^{} \gtrsim \rho_\Sigma^{}\,,~~M_\Sigma^{} \gg \langle\xi\rangle\,.
\end{eqnarray}
Accordingly, the three fermion singlets $S_{R}^{}$ can acquire a Majorana mass term besides their mixing with the heavy fermion singlets $X_{R}^{}$, i.e.
\begin{eqnarray}
\mathcal{L}&\supset&- \frac{1}{2}f_\Sigma^{} \langle \Sigma\rangle \bar{S}_R^{c} S_R^{} - g_X^{}\langle\xi\rangle \bar{S}_R^{} X_R^{c}+\textrm{H.c.}\,.
\end{eqnarray}
In addition, the three fermion singlets $S_{R}^{}$ can have a mass term with the three right-handed neutrinos $\nu_{R}^{}$ when the Higgs singlet $\chi$ drives the $U(1)_{B-L}^{}$ symmetry breaking,
\begin{eqnarray}
\mathcal{L}&\supset& -m_N^{} \bar{\nu}_{R}^{} S_R^{c} +\textrm{H.c.}^{} ~~\textrm{with}~~m_N^{} = f\langle\chi\rangle\,.
\end{eqnarray}
Furthermore, the right-handed neutrinos $\nu_R^{}$ can obtain the usual Dirac masses with the left-handed neutrinos $\nu_L^{}$ when the Higgs doublet $\phi$ acquires its VEV for the electroweak symmetry breaking, 
\begin{eqnarray}
\mathcal{L}&\supset& -m_D^{} \bar{\nu}_{L}^{} \nu_R^{} +\textrm{H.c.}^{} ~~\textrm{with}~~m_D^{} = y\langle\phi\rangle\,.
\end{eqnarray}
In conlusion, the four types of neutral fermions $\nu_L^{}$, $\nu_R^{}$, $S_R^{}$ and $X_R^{}$ have the masses as below, 
\begin{eqnarray}
\!\!\!\!\!\!\mathcal{L}_m^{}\!\!&=&\!\! -\frac{1}{2}\!\left[\bar{\nu}_L^{}\,\bar{\nu}_R^c\, \bar{S}_R^c \,\bar{X}_R^c\right] \!\!\!\left[\begin{array}{cccc} 0& m_D^{}& 0&0\\
[0.75mm]
m_D^T& 0 & m_N^{\ast}&0\\
[0.75mm]
0& m_N^\dagger & g_\Sigma \langle\Sigma\rangle& g_X^{\ast} \langle\xi\rangle\\
[0.75mm]
0&0& g_X^{\dagger} \langle\xi\rangle & M_X^{}\end{array}\right]\!\!\!\! \left[ \begin{array}{c} \nu_L^c\\
[1mm]
\nu_R^{}\\
[1mm]
S_R\\
[1mm]
X_R^{}\end{array}\right] \nonumber\\
\!\!\!\!\!\!\!\!&&\!\!+\textrm{H.c.}\,.  
\end{eqnarray}

The above symmetric mass matrix can be block diagonalized since the element $M_X^{}$ is expected much larger than the other elements, i.e.  
\begin{eqnarray}
\mathcal{L}_m^{}\!\!&\simeq&\!\! -\frac{1}{2}\!\left[\bar{\nu}_L^{}~\bar{\nu}_R^c~ \bar{S}_R^c ~\bar{X}_R^c\right] \!\!\!\left[\begin{array}{cccc} 0 & m_D^{} & 0&0\\
[0.75mm]
m_D^T& 0 & m_N^{\ast}&0\\
[0.75mm]
0& m_N^\dagger & \mu_S^{}& 0\\
[0.75mm]
0&0& 0 & M_X^{}\end{array}\right]\!\!\!\! \left[ \begin{array}{c} \nu_L^c\\
[1mm]
\nu_R^{}\\
[1mm]
S_R\\
[1mm]
X_R^{}\end{array}\right] \nonumber\\
\!\!&&\!\!+\textrm{H.c.}~~\textrm{with}\nonumber\\
\!\!&& \!\! \mu_S^{} = g_{\Sigma}^{}\langle\Sigma\rangle-g_{X}^{\ast} \frac{ \langle\xi\rangle^2_{}}{M_{X}^{}} g_{X}^\dagger =\mu^T_S\,.  
\end{eqnarray}
Clearly, as the Majorana fermions $X$ and the Higgs singlets $\Sigma$ are assumed heavy enough, the Majorana masses $\mu_S^{}$ can be highly suppressed in a natural way. Remarkably, this Majorana mass generation is similar to the conventional type-I \cite{minkowski1977} and type-II \cite{mw1980} seesaw mechanisms, as shown in Fig. \ref{numass}.

Subsequently, in the limiting case,
\begin{eqnarray}
m_N^{}\gg \mu_S^{}\,,~m_D^{}\,,
\end{eqnarray}
the fermion singlets $S_R^{}$ and the right-handed neutrinos $\nu_R^{}$ can form three quasi-Dirac particles, 
\begin{eqnarray}
N=\nu_R^{}+S_R^c\,,
\end{eqnarray}
while the left-handed neutrinos $\nu_L^{}$ can obtain their small Majorana masses,
\begin{eqnarray}
\label{inverse}
\mathcal{L}_m^{}\!\!&\simeq&\!\! -\frac{1}{2}\!\left[\bar{\nu}_L^{}~\bar{\nu}_R^c~ \bar{S}_R^c ~\bar{X}_R^c\right] \!\!\!\left[\begin{array}{cccc} m_\nu^{} &0 & 0&0\\
[0.75mm]
0 & 0 & m_N^{\ast}&0\\
[0.75mm]
0& m_N^\dagger & \mu_S^{} & 0\\
[0.75mm]
0&0&0 & M_X^{}\end{array}\right]\!\!\!\! \left[ \begin{array}{c} \nu_L^c\\
[1mm]
\nu_R^{}\\
[1mm]
S_R\\
[1mm]
X_R^{}\end{array}\right] \nonumber\\
\!\!&&\!\!+\textrm{H.c.}\nonumber\\
\!\!&\simeq & \!\!  -\frac{1}{2}m_\nu^{} \bar{\nu}_{L}^{} \nu_L^{c} - m_N^{} \bar{\nu}_R^{} S_R^{} - \frac{1}{2} M_X^{} \bar{X}_R^{c} X_R^{} +\textrm{H.c.}\nonumber\\
\!\!&&\!\!\textrm{with}~~m_\nu^{} =   m_D^{} \frac{1}{m_N^{\dagger}}\mu_S^{} \frac{1}{m_N^{\ast}} m_D^T\,.
\end{eqnarray}
Here the formula of the small neutrino masses $m_\nu^{}$ is known as the inverse seesaw \cite{mv1986}.

\section{Heavy Higgs and fermion singlet decays}

\begin{figure*}
\vspace{7cm} \epsfig{file=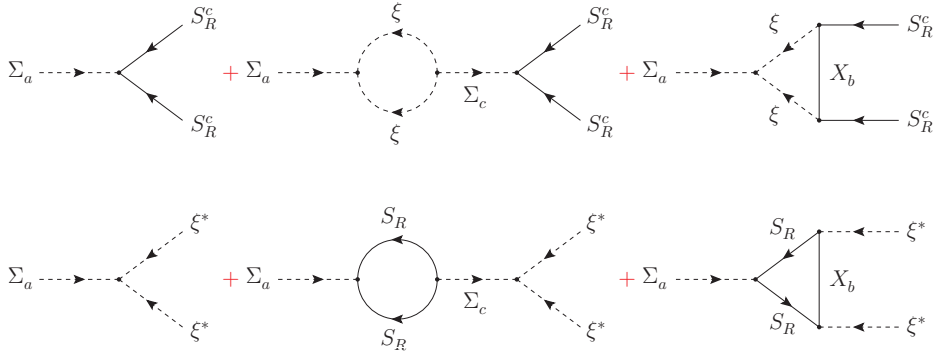, bbllx=7.5cm, bblly=6.0cm,
bburx=17.5cm, bbury=16cm, width=6cm, height=6cm, angle=0,
clip=0} \vspace{-6cm} \caption{\label{sdecay} The heavy Higgs singlet decays.}
\end{figure*}

As shown in Fig. \ref{sdecay}, there are two decay modes of the heavy Higgs singlet $\Sigma_a^{}$, i.e.
\begin{eqnarray}
\Sigma_a^{} \rightarrow  S_R^{c}+S_R^{c}\,,~~ \Sigma_a^{} \rightarrow \xi^\ast_{}+\xi^\ast_{} \,.
\end{eqnarray}
If the CP is not conserved, we can expect a CP asymmetry in the above decays,
\begin{eqnarray}
\varepsilon_{\Sigma_a^{}}^{}&=&2 \frac{\Gamma(\Sigma_a^{} \rightarrow S_R^{c}+S_R^{c}  )-\Gamma( \Sigma_a^{\ast} \rightarrow  S_R^{}+S_R^{}  )}{\Gamma_{\Sigma_a^{}}^{}}\nonumber\\
&=&2\frac{\Gamma( \Sigma^\ast_{a} \rightarrow \xi+\xi)-\Gamma( \Sigma_a^{} \rightarrow \xi^\ast_{}+\xi^\ast_{})}{\Gamma_{\Sigma_a^{}}^{}}\neq 0\,,
\end{eqnarray}
where $\Gamma_{\Sigma_a^{}}^{}$ is the total decay width,
\begin{eqnarray}
\Gamma_{\Sigma_a^{}}^{}&=&\Gamma(\Sigma_a^{} \rightarrow S_R^{c}+S_R^{c})  + \Gamma( \Sigma_a^{} \rightarrow \xi^\ast_{}+\xi^\ast_{})\nonumber\\
&=&\Gamma( \Sigma_a^{\ast} \rightarrow  S_R^{}+S_R^{}  )+\Gamma( \Sigma^\ast_{a} \rightarrow \xi+\xi)\,.
\end{eqnarray}
We can calculate the decay width at tree level and the CP asymmetry at one-loop order,
\begin{eqnarray}
\Gamma_{\Sigma_a^{}}^{}&=&\frac{1}{8\pi}\left[\textrm{Tr}\left(g_{\Sigma_a^{}}^\dagger g_{\Sigma_a^{}}^{}\right)+\frac{\rho_{\Sigma_a^{}}^2}{M_{\Sigma_b^{}}^2}\right]M_{\Sigma_a^{}}^{}\,,\\
\varepsilon_{\Sigma_a^{}}^{}&=& -\frac{1}{\pi}\left\{\sum_{c\neq a}^{}\frac{\textrm{Im}\left[\textrm{Tr}\left(g_{\Sigma_a^{}}^\dagger g_{\Sigma_c^{}}^{}\right)\right]}{\textrm{Tr}\left(g^\dagger_{\Sigma_a^{}} g_{\Sigma_a^{}}^{}\right)+ \frac{\rho_{\Sigma_a^{}}^2}{M_{\Sigma_a^{}}^{2}}}\frac{\rho_{\Sigma_a^{}}^{}\rho_{\Sigma_c^{}}^{}}{M_{\Sigma_c^{}}^2-M_{\Sigma_a^{}}^2}\right.\nonumber\\
&&\left.+\sum_{b}^{}\frac{\textrm{Im}\left[\left(g_X^{\dagger}g_{\Sigma_a^{}}^\dagger g_{X}^{\ast}\right)_{bb}^{}\right]}{\left(g^\dagger_{\Sigma_a^{}}g_{\Sigma_a^{}}^{}\right)_{aa}^{}+ \frac{\rho_{\Sigma_a^{}}^2}{M_{\Sigma_a^{}}^{2}}}\right.\nonumber\\
&&\left.\times \frac{\rho_{\Sigma_a^{}}^{}M_{X_b^{}}^{}}{M_{\Sigma_a^{}}^2}
\ln\left(1+\frac{M_{\Sigma_a^{}}^2}{M_{X_b^{}}^2}\right)\right\}\,.
\end{eqnarray}
Here the first term in the CP asymmetry is the self-energy correction mediated by the heavy Higgs singlet(s) while the second term is the vertex correction mediated by the heavy fermion singlet(s). A nonzero CP asymmetry $\varepsilon_{\Sigma_a^{}}^{}$ needs at least two heavy Higgs singlets $\Sigma$, or at least one heavy Higgs singlet $\Sigma$ and at least one heavy fermion singlet $X$.

\begin{figure*}
\vspace{6.5cm} \epsfig{file=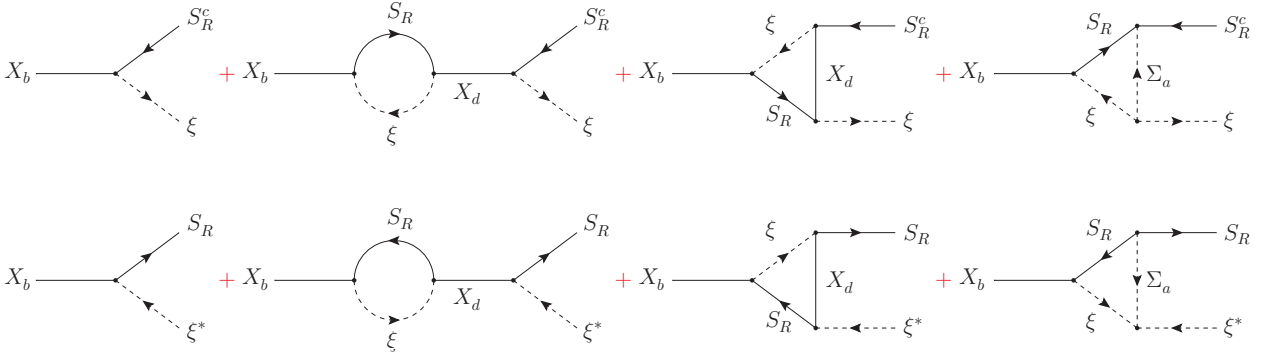, bbllx=11.5cm, bblly=6.0cm,
bburx=21.5cm, bbury=16cm, width=6cm, height=6cm, angle=0,
clip=0} \vspace{-6cm} \caption{\label{fdecay} The heavy fermion singlet decays.}
\end{figure*}

We also consider the decays of the heavy fermion singlet $X_b^{}$. The decay modes are 
\begin{eqnarray}
X_b^{} \rightarrow S_R^{c}+\xi\,,~~ X_b^{} \rightarrow S_R^{}+ \xi^\ast_{} \,.
\end{eqnarray}
The relevant diagrams are shown in Fig. \ref{fdecay}. The decay width and the CP asymmetry can be calculated at tree level and one-loop order, respectively,  
\begin{eqnarray}
\Gamma_{X_b^{}}^{}&=& \Gamma(X_b^{} \rightarrow S_R^{c}+\xi)+\Gamma( X_b^{} \rightarrow S_R^{}+ \xi^\ast_{})\nonumber\\
&=&\frac{1}{16\pi} \left(g_X^\dagger g_X^{}\right)_{bb}^{}M_{X_b^{}}^{}\\
[2mm]
\varepsilon_{X_b^{}}^{}&=& \frac{\Gamma(X_b^{} \rightarrow S_R^c+\xi)- \Gamma(X_b^{} \rightarrow S_R^{}+\xi^\ast_{})}{\Gamma_{X_b^{}}^{}}\nonumber\\
&=&-\frac{1}{16\pi}\left\{\sum_{d\neq b}^{}\frac{\textrm{Im}\left\{\left[\left(g_{X}^T g_{X}^{\ast}\right)_{bd}\right]^2_{}\right\}}{\left(g_X^\dagger g_X^{}\right)_{bb}^{}}\right.\nonumber\\
&&\left.\times \left\{\frac{M_{X_b^{}}^{}M_{X_d^{}}^{}}{M_{X_d^{}}^2- M_{X_b^{}}^2}\right.\right.\nonumber\\
&&\left.+\frac{2M_{X_d^{}}^{}}{M_{X_b^{}}^{}}\left[1-\left(1+\frac{M_{X_d^{}}^{2}}{M_{X_b^{}}^{2}}\right)\ln\left(1+\frac{M_{X_b^{}}^{2}}{M_{X_d^{}}^{2}}\right)\right]\right\}\nonumber\\
&&\left.+\sum_{a}^{}\frac{\textrm{Im}\left[\left(g_{X}^T g_{\Sigma_a^{}}^{} g_{X}^\ast\right)_{bb}^{}\right]}{\left(g_X^\dagger g_X^{}\right)_{bb}^{}}\right.\nonumber\\
&&\left.\times \frac{4\rho_{\Sigma_a^{}}^{}}{M_{X_b^{}}^{}}\left[1-\frac{M_{\Sigma_a^{}}^2}{M_{X_b^{}}^2}\ln\left(1+\frac{M_{X_b^{}}^2}{M_{\Sigma_a^{}}^2}\right)\right]\right\}\,.
\end{eqnarray}
Here the first term in the CP asymmetry is the self-energy and vertex corrections mediated by the heavy fermion singlet(s), while the second term is the vertex corrections mediated by the heavy Higgs singlet(s). A nonzero CP asymmetry $\varepsilon_{X_b^{}}^{}$ needs at least two heavy fermion singlets $X$, or at least one heavy fermion singlet $X$ and at least one heavy Higgs doublet $\Sigma$.

After the heavy Higgs singlets $\Sigma_a^{}$ and the heavy fermion singlets $X_b^{}$ go out of equilibrium, their decays can generate an $X$-asymmetry $X_S^{}$ stored in the fermion singlets $S_R^{}$. For demonstration, we can simply assume the lightest heavy Higgs singlet $\Sigma_1^{}$ or the lightest heavy fermion singlet $X_1^{}$ to be much lighter than the other heavy Higgs and fermion singlets. The $X$-asymmetry $X_S^{}$ then should mainly come from the $\Sigma_1^{}$ or $X_1^{}$ decays, i.e.
\begin{eqnarray}
\label{xasymmetry}
X_S^{}&=& \varepsilon_{\Sigma_1^{}/X_1^{}}^{}\left(\frac{n^{eq}_{\Sigma_1^{}/X_1^{}} }{s}\right)\left|_{T=T_D^{}}^{}\right.,
\end{eqnarray}
where the symbols $n^{eq}_{\Sigma_1^{}/X_1^{}}$ and $T_D^{}$ respectively are the equilibrium number density and the decoupled temperature of the lightest heavy Higgs or fermion singlets, while the character $s$ is the entropy density of the universe \cite{kt1990}. In this case, the CP asymmetries $\varepsilon_{\Sigma_1^{}/X_1^{}}^{}$ can be simplified by 
\begin{eqnarray}
\varepsilon_{\Sigma_1^{}}^{}&\simeq & \frac{1}{\pi}\frac{\textrm{Im}\left[\textrm{Tr}\left(g_{\Sigma_1^{}}^\dagger \mu_S^{} \right)\right] \rho_{\Sigma_1^{}}^{}}{\left[\textrm{Tr}\left(g^\dagger_{\Sigma_1^{}} g_{\Sigma_1^{}}^{}\right)+ \frac{\rho_{\Sigma_1^{}}^2}{M_{\Sigma_1^{}}^{2}} \right] \langle\xi\rangle^2_{} }\nonumber\\
&\leq&  \frac{1}{\pi}\frac{\textrm{Im}\left[\textrm{Tr}\left(g_{\Sigma_1^{}}^\dagger \mu_S^{} \right)\right] \rho_{\Sigma_1^{}}^{}}{2\sqrt{\textrm{Tr}\left(g^\dagger_{\Sigma_1^{}} g_{\Sigma_1^{}}^{}\right)\frac{\rho_{\Sigma_1^{}}^2}{M_{\Sigma_1^{}}^{2}} } \langle\xi\rangle^2_{} }\nonumber\\
&\lesssim&   \frac{1}{2\pi}\frac{ \mu_{\textrm{max}}^{} M_{\Sigma_1^{}}^{}}{ \langle\xi\rangle^2_{} }\,,\nonumber\\
[2mm]
\varepsilon_{X_1^{}}^{}&\simeq & \frac{1}{8\pi}\frac{\textrm{Im}\left[\left(g_{X}^T \mu_S^{} g_{X}^{}\right)_{11}\right] M_{X_1^{}}^{}}{\left(g_X^\dagger g_X^{}\right)_{11}^{}\langle\xi\rangle^2_{}}\nonumber\\
&\lesssim& \frac{1}{8\pi}\frac{\mu_{\textrm{max}}^{} M_{X_1^{}}^{}}{\langle\xi\rangle^2_{}} \,,
\end{eqnarray}
with $\mu_{\textrm{max}}^{} $ being the largest eigenvalue of the Majorana mass matrix $\mu_S^{}$.

Note the $S_R^{}-S_R^c$ oscillation induced by the Majorana masses $\mu_S^{}$ will tend to wash out the $X_S^{}$ asymmetry. However, this oscillation will go into equilibrium at a very low temperature, where the sphalerons are no longer active, i.e
\begin{eqnarray}
&&\left[\Gamma_{S_R^{}-S_R^c}^{}>H(T)\right]\left|_{T\ll  100\,\textrm{GeV}}^{}\right.\nonumber\\
&&\textrm{with}~~\Gamma_{S_R^{}-S_R^c}^{}\sim\left\{\begin{array}{cc} \frac{\mu_S^2}{T}&\textrm{for}~~T>m_N^{}\,,\\
[2mm]
\frac{\mu_S^2}{m_N^{}}&\textrm{for}~~T<m_N^{}\,.\end{array}\right.
\end{eqnarray}
Here $H(T)$ is the Hubble constant,
\begin{eqnarray} 
H(T)=\left[\frac{8\pi^{3}_{}g_{\ast}^{}(T)}{90}\right]^{\frac{1}{2}}_{}\frac{T^2_{}}{M_{\textrm{Pl}}^{}}\,,
\end{eqnarray}
with $M_{\textrm{Pl}}^{}\simeq 1.22\times 10^{19}_{}\,\textrm{GeV}$ being the Planck mass and $g_{\ast}^{}(T)$ being the relativistic degrees of freedom.

\section{Baryon asymmetry}

The $X$-asymmetry $X_S^{}$ stored in the fermion singlets $S_R^{}$ will lead to an $X$-asymmetry stored in the SM lepton doublets $l_L^{}$ because of the related Yukawa interactions in Eq. (\ref{lag}). The sphaleron processes then can partially transfer this SM $X$-asymmetry to a baryon asymmetry. 

We now analysize the chemical potentials \cite{ht1990} to discuss the details of these conversions. For this purpose, we denote $\mu_{q}^{}$, $\mu_d^{}$, $\mu_u^{}$, $\mu_l^{}$, $\mu_e^{}$, $\mu_\nu^{}$, $\mu_S^{}$, $\mu_\phi^{}$ and $\mu_\chi^{}$ for the chemical potentials of the fields $q_{L}^{}$, $d_{R}^{}$, $u_{R}^{}$, $l_L^{}$, $e_R^{}$, $\nu_R^{}$, $S_R^{}$, $\phi$ and $\chi$. We then can consider the chemical potentials in three phases,
\begin{itemize}
\item phase-I: before the $U(1)_{B-L}^{}$ symmetry breaking.
\item phase-II: after the $U(1)_{B-L}^{}$ symmetry breaking.
\item phase-II: after the quasi-Dirac fermions $N$ decays.
\end{itemize}

In phase-I, the SM Yukawa interactions are in equilibrium and hence yield,
\begin{eqnarray}
\label{chemical1}
-\mu_{q}^{}+\mu_{d}^{}-\mu_{\phi}^{}&=&0\,,\\
\label{chemical2}
 -\mu_{q}^{}+\mu_{u}^{}+\mu_{\phi}^{}&=&0\,,\\
 \label{chemical3} 
 -\mu_{l}^{}+\mu_{e}^{}-\mu_{\phi}^{}&=&0\,,
\end{eqnarray}
the fast sphalerons constrain,
\begin{eqnarray}
\label{chemical4}
3\mu_{q}^{}+\mu_{l}^{}&=&0\,,
\end{eqnarray}
while the neutral hypercharge in the universe requires,
\begin{eqnarray}
\label{chemical5}
3\left( \mu_{q}^{} -\mu_{d}^{}+2\mu_{u}^{}-\mu_{l}^{} -\mu_{e}^{}\right)-2\mu_{\phi}^{} =0\,.
\end{eqnarray}
In addition, the Yukawa interactions involving the right-handed neutrinos are also in equilibrium. This means
\begin{eqnarray}
\label{chemical6}
-\mu_{l}^{}+\mu_{\nu}^{}+\mu_{\phi}^{}&=&0\,,\\
\label{chemical6}
-\mu_S^{}-\mu_\nu^{}+\mu_\chi^{}&=&0\,.
\end{eqnarray}
Furthermore, the total $U(1)_{B-L}^{}$ charge should be zero,
\begin{eqnarray}
\label{chemical7}
3\left( 2\mu_{q}^{} +\mu_{d}^{}+\mu_{u}^{}\right) - 3 \left(2\mu_l^{}+\mu_{e}^{}+\mu_\nu^{}\right)-2\mu_{\chi}^{}=0\,.
\end{eqnarray}
In the above Eqs. (\ref{chemical1}-\ref{chemical7}), we have identified the chemical potentials of the different-generation fermions because the Yukawa interactions establish an equilibrium between the different generations. By solving Eqs. (\ref{chemical1}-\ref{chemical7}), we can determine the chemical potentials in phase-I as below,
\begin{eqnarray}
&&\mu_\phi^{I}= -\frac{4}{7}\mu_l^{I} \,,~~\mu_q^{I}= -\frac{1}{3} \mu_l^{I}\,, ~~\mu_d^{I}= -\frac{19}{21}\mu_l^{I}\,,\nonumber\\
&&\mu_u^{I}= \frac{5}{21}\mu_l^{I}\,,~~\mu_e^{I} = \frac{3}{7}\mu_l^{I} \,,~~\mu_{\nu}^{I}=\frac{11}{7}\mu_l^{I} \,,\nonumber\\
&&\mu_S^{I}=-\frac{67}{7}\mu_l^I\,,~~\mu_\chi^{}=-8\mu_l^{I}\,.
\end{eqnarray}

Now the global baryon number can be given by
\begin{eqnarray}
B^I_{}= 3(2\mu_q^{I} + \mu_d^{I} + \mu_u^{I}) = -4 \mu_l^{I}\,.
\end{eqnarray}
As for the global $X$-number, it should be 
\begin{eqnarray}
X^{I}_{}= X_{q_L^{}+d_R^{}+u_R^{}}^{I}  + X_{l_L^{}+e_R^{}}^{I} + X_{\nu_R^{}}^{I}+X_{S_R^{}}^{I}\,,
\end{eqnarray}
with $X_{q_L^{}+d_R^{}+u_R^{}}^{I} $, $X_{l_L^{}+e_R^{}}^{I}$, $X_{\nu_R}^{I}$ and $X_{S_R^{}}^{I}$ being the global $X$-number in the SM quarks, the SM leptons, the right-handed neutrinos $\nu_R^{}$, and the fermion singlets $S_R^{}$, respectively, 
\begin{eqnarray}
X_{q_L^{}+d_R^{}+u_R^{}}^{I}&=& -3(2\mu_q^{I} + \mu_d^{I} + \mu_u^{I}) = 4 \mu_l^{I}\,,\nonumber\\
X_{l_L^{}+e_R^{}}^{I}&=& 3(2\mu_l^{I} + \mu_e^{I} ) = \frac{51}{7}\mu_l^{I} \,,\nonumber\\
X_{\nu_R^{}}^{I}&=&3 \mu_\nu^{I} = \frac{33}{7}\mu_l^{I}\,,\nonumber\\
X_{S_R^{}}^{I}&=&-3\mu_{S}^{I}= \frac{201}{7}\mu_l^{I}\,.
\end{eqnarray}
The global $X$-number then can be computed by  
\begin{eqnarray}
\label{bl1}
X^{I}_{}&=& X_{q_L^{}+d_R^{}+u_R^{}}^{I} +  X_{l_L^{}+e_R^{}}^{I}  + L_{\nu_R^{}}^{I}+L_{S_R^{}}^{I} \nonumber\\
&=& \frac{313}{7}\mu_l^{I} \,.
\end{eqnarray}

The global $X$-number (\ref{bl1}) should be conserved after it is produced from the lightest heavy Higgs or fermion singlet decays. Therefore, we can read
\begin{eqnarray}
X^{I}_{}&=& X^{i}_{}=  X^{}_S   \nonumber\\
&=& \varepsilon_{\Sigma_1^{}/X_1^{}}^{}\left(\frac{n^{eq}_{\Sigma_1^{}/X_1^{}} }{s}\right)\left|_{T=T_D^{}}^{}\right. ,
\end{eqnarray}
where $X^i_{}=X^{}_S$ is the initial $X$-number from the lightest heavy Higgs or fermion singlet decays. So we eventually can derive
\begin{eqnarray}
X^{I}_{q_L^{}+d_R^{} + u_R^{}}&=&\frac{28}{313}  X^{I}_{}=\frac{28}{313} X^{}_{S}\,,\nonumber\\
X^{I}_{l_L^{}+e_R^{}}&=&\frac{51}{313}X^{I}_{}=\frac{51}{313} X^{}_{S}\,,\nonumber\\
X_{\nu_R^{}}^{I}&=&\frac{33}{313}X^{I}_{}=\frac{33}{313} X^{}_{S}\,,\nonumber\\
X_{S_R^{}}^{I}&=&\frac{201}{313} X^{I}_{}=\frac{201}{313} X^{}_{S}\,,\nonumber\\
B^{I}_{}&=&- X^{I}_{q_L^{}+d_R^{} + u_R^{}}= -\frac{28}{313} X^{}_{S}\,.
\end{eqnarray}
The $U(1)_{B-L}^{}$ gauge symmetry might be broken after the electroweak symmetry breaking if its gauge coupling is small enough to escape from the experimental limits. In this case, the final baryon asymmetry can be determined at this moment, i.e.  
\begin{eqnarray}
\label{bauf1}
B_f^{}&=&B^{I}_{}=-\frac{28}{313}X^{}_{S}\nonumber\\
&=&-\frac{28}{313}\varepsilon_{\Sigma_1^{}/X_1^{}}^{}\left(\frac{n^{eq}_{\Sigma_1^{}/X_1^{}} }{s}\right)\left|_{T=T_D^{}}^{}\right. .
\end{eqnarray}

Usually the $U(1)_{B-L}^{}$ gauge symmetry should be broken before the electroweak symmetry breaking as its gauge coupling is not chosen to be very small. We thus consider the phase-II, where Eq. (\ref{chemical7}) should be removed while Eq. (\ref{chemical6}) should be modified,
\begin{eqnarray}
\label{chemical8}
-\mu_S^{}-\mu_\nu^{}=0\,.
\end{eqnarray}  
The chemical potentials of the relativistic particles now are given by 
\begin{eqnarray}
&&\mu_\phi^{II}=-\frac{4}{7}\mu_l^{II}\,,~~\mu_q^{II}=-\frac{1}{3}\mu_l^{II}\,,~~\mu_d^{II}=-\frac{19}{21}\mu_l^{II}\,,\nonumber\\
&&\mu_u^{II}=\frac{5}{21}\mu_l^{II}\,,~~\mu_{e}^{II}=\frac{3}{7}\mu_l^{II}\,,~~\mu_{\nu}^{II}=\frac{11}{7}\mu_l^{II}\,,\nonumber\\
&&\mu_S^{II}=-\frac{11}{7}\mu_l^{II}\,.
\end{eqnarray}
At this stage, the global baryon number in the SM quarks and the global $X$-numbers in the SM quarks, the SM leptons, the right-handed neutrinos $\nu_R^{}$, and the fermion singlets $S_R^{}$ should be 
\begin{eqnarray}
X^{II}_{q_L^{}+d_R^{}+u_R^{}}&=&-3(2\mu_q^{II}+\mu_d^{II}+\mu_u^{II})=4\mu_{l}^{II}\,,\nonumber\\
X_{l_L^{}+e_R^{}}^{II}&=& 3(2\mu_l^{II} + \mu_e^{II})= \frac{51}{7}\mu_l^{II} \,,\nonumber\\
X_{\nu_R^{}}^{II}&=&3 \mu_\nu^{II} = \frac{33}{7}\mu_l^{II}\,,\nonumber\\
X_{S_R^{}}^{II}&=&-3\mu_{S_R^{}}^{II}= \frac{33}{7}\mu_l^{II}\,,\nonumber\\
B^{II}_{}&=&-X^{II}_{q_L^{}+d_R^{}+u_R^{}}=-4\mu_{l}^{II}\,.
\end{eqnarray}
The conserved $X$-number in the phase-II thus becomes to be 
\begin{eqnarray}
X^{II}_{}&=& X_{q_L^{}+d_R^{}+u_R^{}}^{II}+X_{l_L^{}+e_R^{}}^{II} + X_{\nu_R^{}}^{II}+X_{S_R^{}}^{II}=\frac{145}{7}\mu_l^{II}\nonumber\\
&=& X^{I}_{}=X_S^{}\,.
\end{eqnarray}
Accordingly, we can read 
\begin{eqnarray}
X^{II}_{q_L^{}+d_R^{}+u_R^{}}&=&-\frac{28}{145}X^{II}_{}=\frac{28}{145}X^{}_{S}\,,\nonumber\\
X^{II}_{l_L^{}+e_R^{}}&=&\frac{51}{145}X^{II}_{}=\frac{51}{145}X^{}_{S}\,,\nonumber\\
X_{\nu_R^{}}^{II}&=&\frac{33}{145}X^{II}_{}=\frac{33}{145}X^{}_{S}\,,\nonumber\\
X_{S_R^{}}^{II}&=&\frac{33}{145}X^{II}_{}=\frac{33}{145}X^{}_{S}\,,\nonumber\\
B^{II}_{}&=& -X^{II}_{q_L^{}+d_R^{}+u_R^{}} =- \frac{28}{145}X^{}_{S}\,.
\end{eqnarray}
Now the right-handed neutrinos $\nu_R^{}$ and the fermion singlets $S_R^{}$ have formed three quasi-Dirac particles $N=\nu_R^{}+S_R^{c}$. If these $N$ fermions keep relativistic before the electroweak symmetry breaking, the final baryon asymmetry can be simply given by  
\begin{eqnarray}
\label{bauf2}
B^f_{}&=&B^{II}_{}=-\frac{28}{145} X_S^{} \nonumber\\
&=&  -\frac{28}{145} \varepsilon_{\Sigma_1^{}/X_1^{}}^{}\left(\frac{n^{eq}_{\Sigma_1^{}/X_1^{}} }{s}\right)\left|_{T=T_D^{}}^{}\right. .
\end{eqnarray}

Alternatively, the $N$ fermions have already become non-relativistic and hence have completely decayed before the electroweak symmetry breaking. In this case, we should consider the phase-III where the chemical potentials of the relativistic freedoms are
\begin{eqnarray}
\label{chemical8}
&&\mu_\phi^{III}=-\frac{4}{7}\mu_l^{III}\,,~~\mu_q^{III}=-\frac{1}{3}\mu_l^{III}\,,~~\mu_d^{III}=-\frac{19}{21}\mu_l^{II}\,,\nonumber\\
&&\mu_u^{III}=\frac{5}{21}\mu_l^{III}\,,~~\mu_{e}^{III}=\frac{3}{7}\mu_l^{III}\,,
\end{eqnarray}
and hence the global baryon number and the global $X$-number in the SM are 
\begin{eqnarray}
 X^{III}_{q_L^{}+d_R^{}+u_R^{}}&=&-3(2\mu_q^{III} + \mu_d^{III} + \mu_u^{III}) = 4 \mu_l^{III}\,,\nonumber\\
 X_{l_L^{}+e_R^{}}^{III} &=&3(2\mu_l^{III} + \mu_e^{III})= \frac{51}{7}\mu_l^{III}\,,\nonumber\\ 
 B^{III}_{}&=&-X^{III}_{q_L^{}+d_R^{}+u_R^{}}=-4\mu_l^{III}\,. 
 \end{eqnarray}
The conversed $X$-number in the phase-III thus should be
 \begin{eqnarray}
X^{III}_{}&=& X^{III}_{q_L^{}+d_R^{}+u_R^{}} +X_{l_L^{}+e_R^{}}^{III}=\frac{79}{7}\mu_l^{III} \nonumber\\
&=& X^{II}_{}=X_S^{}\,.
\end{eqnarray}
We hence can obtain
\begin{eqnarray}
X^{III}_{q_L^{}+d_R^{}+u_R^{}}&=&-\frac{28}{79}X^{II}_{}=\frac{28}{79}X^{}_{S}\,,\nonumber\\
X^{III}_{l_L^{}+e_R^{}}&=&\frac{51}{79}X^{II}_{}=\frac{51}{79}X^{}_{S}\,,\nonumber\\
B^{III}_{}&=& -X^{III}_{q_L^{}+d_R^{}+u_R^{}} =- \frac{28}{79}X^{}_{S}\,.
\end{eqnarray}
The final baryon number then can be given by 
\begin{eqnarray}
\label{bauf3}
B^f_{}&=&B^{III}_{}= -\frac{28}{79} X_S^{}  \nonumber\\
&=&-\frac{28}{79} \varepsilon_{\Sigma_1^{}/X_1^{}}^{}\left(\frac{n^{eq}_{\Sigma_1^{}/X_1^{}} }{s}\right)\left|_{T=T_D^{}}^{}\right. .
\end{eqnarray}

\section{Numerical example}

We can solve the related Boltzmann equations to determine the $X$-asymmetry (\ref{xasymmetry}) and then the final baryon asymmetry (\ref{bauf1}), (\ref{bauf2}) or (\ref{bauf3}) by inputting the masses and couplings of the heavy Higgs or fermion singlets $\Sigma_1^{}/X_1^{}$. For simplicity, we just consider the weak washout region where the lightest heavy Higgs or fermion singlet decays match the condition, 
\begin{eqnarray}
\label{weak}
\left.\left[\Gamma_{1}^{}< H(T)=\left[\frac{8\pi^{3}_{}g_{\ast}^{}(T)}{90}\right]^{\frac{1}{2}}_{}\frac{T^2_{}}{M_{\textrm{Pl}}^{}}\right]\right|_{T=M_{\Sigma_1^{}/X_1^{}}}^{}\,,
\end{eqnarray}
so that the final baryon asymmetry can be approximately described by    
\begin{eqnarray}
\label{bauf4}
B^f_{}\sim c \frac{\varepsilon_1^{}}{g_\ast^{}}~~\textrm{with}~~c=-\frac{28}{79}~\textrm{or}~-\frac{28}{145}~\textrm{or}~-\frac{28}{313}\,.
\end{eqnarray}
Here $H(T)$ is the Hubble constant with $M_{\textrm{Pl}}^{}\simeq 1.22\times 10^{19}_{}\,\textrm{GeV}$ being the Planck mass and $g_{\ast}^{}(T)=122.25$ being the relativistic degrees of freedom (the SM fields plus the three right-handed neutrinos $\nu_{R}^{}$, the three gauge-singlet fermions $S_R^{}$, the $U(1)_X^{}$ Higgs singlet $\xi$, the $U(1)_{B-L}^{}$ Higgs singlet $\chi$ and the $U(1)_{B-L}^{}$ gauge boson.).

As an example, we choose 
\begin{eqnarray}
\langle\xi\rangle= \mathcal{O}(100\,\textrm{TeV})\,,
\end{eqnarray}
and then take
\begin{eqnarray}
\!\!\!\!\!\!\!\!\!\!\!\!&&M_{\Sigma_1^{}}^{}=10^{14}_{}\,\textrm{GeV}\,,~\rho_{\Sigma_1^{}}^{}=10^{12}_{}\,\textrm{GeV}\,, ~(g_{\Sigma_1^{}}^{})_{ij}^{}= \mathcal{O}(0.01)\,; \nonumber\\
\!\!\!\!\!\!\!\!\!\!\!\!&&\textrm{or}~~M_{X_1^{}}^{}=10^{14}_{}\,\textrm{GeV}\,,~(g_{X}^{})_{i1}^{} = \mathcal{O}(0.01)\,.
\end{eqnarray}
With these inputting, we can have
\begin{eqnarray}
\label{output1}
\mu_S^{}= \mathcal{O}(10\,\textrm{eV})\,,
\end{eqnarray}
as well as 
\begin{eqnarray}
\frac{\Gamma_{\Sigma_1^{}/X_1^{}}^{}}{H(T)}\left|_{T=M_{\Sigma_1^{}/X_1^{}}^{}}^{}\right.=\mathcal{O}(0.1)\,,~~\varepsilon_{\Sigma_1^{}/X_1^{}}^{\textrm{max}}=\mathcal{O}( 10^{-5}) \,.
\end{eqnarray}
So, the final baryon asymmetry (\ref{bauf4}) can arrive at the observed value $B^f_{}\sim 10^{-10}_{}$.

We further take 
\begin{eqnarray}
\!\!\!\!\!\!\!\!\!\!\!\!&&\langle\chi\rangle= \mathcal{O}(10\,\textrm{TeV})\,, ~f\sim \mathcal{O}(10^{-3}_{}-0.1)\,,~y\sim \mathcal{O}(10^{-2}_{}-1)\,,\nonumber\\
\!\!\!\!\!\!\!\!\!\!\!\!&&
\end{eqnarray}
to give  
\begin{eqnarray}
\label{output2}
\!\!m_N^{} \sim \mathcal{O}(10-1000\,\textrm{GeV})\,, ~~m_D^{}\sim \mathcal{O}(1-100\,\textrm{GeV})\,.
\end{eqnarray}
By inserting the outputs (\ref{output1}) and (\ref{output2}) into the inverse seesaw (\ref{inverse}), we can obtain the neutrino masses $m_\nu^{}=\mathcal{O}(0.1\,\textrm{eV})$.

For the above parameter choice, we also check the $S_R^{}-S_R^c$ oscillation induced by the Majorana masses $\mu_S^{}$. We find this oscillation can not go into equilibrium before the electroweak symmetry breaking and hence it will not affect the production of the baryon asymmetry.

\section{Conclusion}

In this paper, we have shown the small Majorana masses of the fermion singlets in the inverse seesaw models can have a common origin with the cosmic baryon asymmetry. Our scenario is based on a $U(1)_{B-L}^{}$ gauge symmetry and a $U(1)_X^{}$ global symmetry. When a gauge-singlet Higgs scalar drives the spontaneous breaking of the $U(1)_X^{}$ symmetry, three gauge-singlet fermions can obtain their small Majorana masses by integrating out some heavy gauge-singlet scalars and/or fermions. These gauge-singlet fermions can also mix with the same number of right-handed neutrinos after the $U(1)_{B-L}^{}$ symmetry is spontaneously broken. Thanks to the Yukawa interactions between the right-handed neutrinos and the standard model, the left-handed neutrinos eventually can obtain a small Majorana mass term through the inverse seesaw mechanism. On the other hand, through their couplings for realizing the inverse seesaw, the heavy gauge-singlet scalars and/or fermions can decay to produce an asymmetry stored in the three gauge-singlet fermions. The sphaleron processes then can partially transfer this asymmetry to a baryon asymmetry because of the sizable Yukawa couplings involving the right-handed neutrinos.

\textbf{Acknowledgement}: This work was supported by the National Natural Science Foundation of China under Grant No. 11675100 and the Recruitment Program for Young Professionals under Grant No. 15Z127060004.

\end{document}